\documentstyle[12pt]{article}
\begin{document}

\title{Correlated mean field Ansatz for the Kondo necklace}
\author{H. Y. Kee\thanks{Permanent address:
Department of Physics Education, Seoul National Unversity, Seoul, 151--742
Korea}\ \ and P. Fazekas\thanks{Permanent address: Research Institute
for Solid State Physics, P.O.B. 49, Budapest 114, H--1525 Hungary}\\
International Centre for Theoretical Physics,\\
P.O. Box 586, I--34100 Trieste, Italy}
\date{}
\maketitle

\vskip 1 cm

\begin{abstract}
We study the ground state phase diagram of the pseudospin model introduced
by Doniach to describe the essential physics of Kondo lattices. We use
variational trial states which augment the usual mean field solution by
incorporating various intersite correlations. A composite spin correlation
describing the antiparallel alignment of fluctuating triplets is found to
be particularly favourable for large Kondo couplings. With this trial state,
the magnetic--to--Kondo transition is suppressed and the strong coupling
ground state is ordered with strongly reduced moments. The relevance of the
findings is discussed.

\end{abstract}
\newpage

\section{The model}

The basic question in the physics of heavy fermion materials is whether
collective spin compensation can be taking place in a periodic array of
localized moments immersed in a conduction electron sea \cite{GS}. In case
yes, we may speak about a collective Kondo effect even though it remains
unclear to which extent the formation of an overall lattice singlet ground
state can be likened to the single--ion Kondo effect. The most intriguing
possibility is that spin compensation may go a very long way before
it is stopped by the ordering of the residual tiny moments \cite{CG}.

As far as spins are concerned, the Kondo effect is just a spin compensation
phenomenon. The emergence of a non--analytic energy scale in the impurity
problem is connected with the existence of a large number of arbitrarily
low--lying electron--hole excitations. Though variational methods indicate
the existence of a lattice--coherence--enhanced Kondo energy scale for the
nearly integral valent (Kondo) regime of the Anderson lattice \cite{RU,Fa},
as well as for the Kondo lattice \cite{SF,FS}, they do not provide a proof
that the ground state energy of the Kondo lattice contains non--analytic
terms. The interplay of spin and charge degrees of freedom in the Kondo
lattice may still prove to be quite different from what one has found for
the impurity problem. It seems desirable to separate, if possible, the
spin compensation aspect from all the other complications of ``true Kondo
physics''.

It has become accepted \cite{SM,LiCo} that the competition of spin
compensation and magnetic ordering can be described with drastically
simplified models
which contain just the spin degrees of freedom. The simplest of these is
the Kondo necklace model introduced by Doniach \cite {Don}. In addition to
the localized
spins ${\vec S}$ of the $f$--electrons we introduce a set of pseudospins
${\vec \tau}$'s which stand for the spin degrees of freedom of the
conduction electrons. The number of $\vec S$--spins is chosen equal to the
number of ${\vec \tau}$ spins which implies that the
possibility of a full spin compensation is a feature of the system. While
it can be argued \cite{SF,FM} that a singlet ground state  may arise at any
band filling, it is more straightforward to associate the model with the
Kondo lattice with a half--filled conduction band so that the number of
pseudospins is equal to the number of conduction electrons. This has the
additional motivation that the corresponding Kondo lattice has an
insulating ground state \cite{Tsu}; the appearance of a charge gap is a
justification for omitting the charge degrees of freedom.

The simplest mean field treatment of the original necklace model
yielded the beautiful result of a ground state phase transition from
a magnetically ordered to a fully spin--compensated state \cite{Don}.
We are by now fully aware that strictly one--dimensional models are
bound to show much subtler behaviour \cite{SM} but extending the model to
the physically more relevant higher dimensions (no longer ``necklaces'' in
the geometrical sense) makes us to expect that the mean field results are
roughly correct.

Our intention here is to improve the mean field
approximation by including short--range correlations. Such an approach is
expected to give sensible improvements over single--site mean field results
in three dimensions where the mean field phase diagram should be
qualitatively correct. However, intermediate steps of our calculation can
be executed free of further approximations in one dimension. Therefore, we
make the algebra (sum over local configurations) for the one--dimensional
case but bear in mind that the character of the results is meant for three
dimensions. (In one dimension, the better estimate of the ground
state energy is still believable but one should not trust the
characterization of the ground state.)

We study $S=1/2$ Kondo necklace models
\begin{equation}
H = J\sum_{i=1}^{L}{\vec S}_{i}\cdot{\vec \tau}_{i} +
W\sum_{i=1}^{L}(\tau_{i}^{x}\tau_{i+1}^{x}+\tau_{i}^{y}\tau_{i+1}^{y}
+\rho\tau_{i}^{z}\tau_{i+1}^{z})
\label{eq:ham}
\end{equation}
The necklace is closed with the periodic boundary condition $L+1\equiv 1$.

Seeking
correspondence with the insulating state of the Kondo lattice model would
lead us to choose an antiferromagnetic Kondo coupling $J>0$, and an
antiferromagnetic intersite pseudospin coupling $W>0$. Actually, in Doniach's
\cite{Don} original pseudospin model, the latter term was chosen to be purely
$x$--$y$--like ($\rho=0$), which gives a good imitation of propagating
degrees of freedom. One should remember, however, that the underlying
fermionic Kondo lattice problem had spin--rotational symmetry, and this has
been lost by postulating the $x$--$y$ form of coupling. We note that
isotropic spin models (or anisotropic models with isotropy as a special case)
have been discussed in the literature with the purpose of
modelling Kondo lattice physics \cite{RS,SM}. In  any case, a more complete
understanding of the spin system makes the study of an extended model
desirable. With this  motivation, the study of (\ref{eq:ham}) with arbitrary
signs of $J$ and $W$, and with a general anisotropy $0\le\rho<\infty$ is
indicated. For reasons of convenience, we concentrate on the case $\rho=0$
but we wish to emphasize that the method used here is equally applicable for
$\rho\ne 0$.

Hamiltonians of the form (\ref{eq:ham}) can be defined for either
$|{\vec S}|>|{\vec \tau}|$, or $|{\vec \tau}|>|{\vec S}|$, or
$|{\vec \tau}|=|{\vec S}|$, corresponding to underscreened, overscreened,
or exactly screened Kondo lattices. Generalizing Doniach's work on the
exactly screened $S=1/2$ model, we have discussed the mean field ground
states of the underscreened \cite{FK1} and overscreened \cite{FK2}, Kondo
necklace models earlier. Here our main interest lies in going beyond the
single--site mean field description, therefore we confine our attention to
the simplest case $|{\vec S}|=1/2$, and $|{\vec \tau}|=1/2$. The Hilbert
space of a lattice site is spanned by the four local basis states
$|S^{z}\tau^{z}\rangle$
\begin{eqnarray}
|1\rangle = |1/2, 1/2\rangle &\ \ \ \ |2\rangle & = |-1/2, 1/2\rangle
\nonumber\\
|3\rangle = |1/2, -1/2\rangle &\ \ \ \ |4\rangle & = |-1/2, -1/2\rangle
\label{eq:bas}
\end{eqnarray}

The number of cases is reduced if we set $\rho=0$. It is wellknown that
the ferromagnetic and antiferromagnetic $S=1/2$ $x$--$y$ models (on
bipartite lattices) are physically identical since they can be
connected by a canonical transformation. A bipartite lattice can be divided
into alternate sublattices $A$ and $B$ so that nearest--neighbour bonds always
connect different sublattices. Then the transformation
\begin{equation}
{\hat U}_{1} = \prod_{j\in B} \exp{(-i\pi \tau_{j}^{z})}
\label{eq:uni1}
\end{equation}
changes the sign of the $x$--$y$ term
\begin{equation}
{\hat U}_{1}\left(W\sum_{i=1}^{L}(\tau_{i}^{x}\tau_{i+1}^{x}+
\tau_{i}^{y}\tau_{i+1}^{y})\right){\hat U}_{1}^{-1} =
-W\sum_{i=1}^{L}(\tau_{i}^{x}\tau_{i+1}^{x}+
\tau_{i}^{y}\tau_{i+1}^{y})
\end{equation}
\\
A similar statement holds for our
model: a $\pi$--rotation about the spin--$z$ axes for both the $S$--, and
$\tau$--spins on sublattice $B$
\begin{equation}
{\hat U}_{2} = \prod_{j\in B} \exp{[-i\pi (S_{j}^{z}+\tau_{j}^{z})]}
\label{eq:uni2}
\end{equation}
changes the sign of the intersite  $\tau$--coupling while leaving the Kondo
term unchanged
\begin{eqnarray}
{\hat U}_{2}\left(J\sum_{i=1}^{L}{\vec S}_{i}\cdot{\vec \tau}_{i} +
W\sum_{i=1}^{L}(\tau_{i}^{x}\tau_{i+1}^{x}+
\tau_{i}^{y}\tau_{i+1}^{y})\right){\hat U}_{2}^{-1}\nonumber \\
= J\sum_{i=1}^{L}{\vec S}_{i}\cdot{\vec \tau}_{i}
 - W\sum_{i=1}^{L}(\tau_{i}^{x}\tau_{i+1}^{x}+
\tau_{i}^{y}\tau_{i+1}^{y})
\label{eq:uni3}
\end{eqnarray}
\\
The second term on the right--hand side can be transformed to a
non--interacting spinless fermion model \cite{LM}: for half--filling, the
ground state energy is $-W/\pi$. This provides a useful comparison for
energy estimates in the weak--$J$ regime.

The inclusion of $\rho\ne 0$ would, of course, make the cases of positive
and negative $W$ genuinely different. Our variational method is, in principle,
just as applicable for $\rho\ne 0$ as for $\rho=0$. Restricting our attention
to the case $\rho=0$, where it is sufficient to consider $W<0$, is motivated
by formal convenience: for ferromagnetic intersite coupling, homogeneous trial
states can be used.

\section{Variational method}

We wish to map out the ground state phase diagram of $H$. For this purpose,
we introduce variational trial states. Since a site $i$ has two spins, the
``true'' spin ${\vec S}_{i}$, and the pseudospin ${\vec \tau}_{i}$ which are
coupled by the Kondo term, we can speak of the internal structure of the
site, which is described as a linear combination of the four possible states
\begin{equation}
|\phi\rangle_{i} = \alpha_{1}|1\rangle_{i} + \alpha_{2}|2\rangle_{i} +
\alpha_{3}|3\rangle_{i} + \alpha_{4}|4\rangle_{i}
\label{eq:phi}
\end{equation}
In the single--site mean field theory, the internal state of the site $i$
is taken to be independent of the instantaneous state of any other site; e.g.,
a translationally invariant state would be described by the product wave
function
\begin{equation}
|\Phi\rangle = \prod_{i=1}^{L} |\phi\rangle_{i} =
\prod_{i=1}^{L}(\alpha_{1}|1\rangle_{i} +
\alpha_{2}|2\rangle_{i} +
\alpha_{3}|3\rangle_{i} + \alpha_{4}|4\rangle_{i})
\label{eq:phii}
\end{equation}
In a ground state with antiferromagnetic long--range order,
$|\phi\rangle_{i}$ can be made sublattice--dependent
\begin{eqnarray}
|\Phi\rangle_{AF} & = & \prod_{i=1}^{L/2} |\phi^{A}\rangle_{2i-1}
|\phi^{B}\rangle_{2i} \nonumber \\[3mm]
& = & \prod_{i=1}^{L/2}  (\alpha_{1}^{A}|1\rangle_{2i-1} +
\alpha_{2}^{A}|2\rangle_{2i-1} +
\alpha_{3}^{A}|3\rangle_{2i-1} + \alpha_{4}^{A}|4\rangle_{2i-1})\cdot
\nonumber \\[3mm]
& &   (\alpha_{1}^{B}|1\rangle_{2i} +
\alpha_{2}^{B}|2\rangle_{2i} +
\alpha_{3}^{B}|3\rangle_{2i} + \alpha_{4}^{B}|4\rangle_{2i})
\label{eq:phiAF}
\end{eqnarray}
but otherwise, the simple product form is retained.

We propose to improve the variational description by including
nearest--neighbour intersite correlations, according to the recipe: if site
$i$ is in the state $|\xi_{i}\rangle_{i}$, and site $i+1$ in the state
$|\xi_{i+1}\rangle_{i+1}$, then the amplitude acquires the additional factor
$O(\xi_{i}, \xi_{i+1})$. For a translationally invariant state, we can write
\begin{equation}
|\Psi\rangle = \prod_{i} {\hat P}_{i,i+1}|\Phi\rangle
\label{eq:psi}
\end{equation}
where the intersite correlator
\begin{equation}
{\hat P}_{i,i+1} = \sum_{\xi_{i}=1}^{4} \sum_{\xi_{i+1}=1}^{4}
|\xi_{i}\rangle_{i}
|\xi_{i+1}\rangle_{i+1} \ O(\xi_{i},\xi_{i+1}) \ _{i+1}\langle\xi_{i+1}|
_{i}\langle\xi_{i}|
\label{eq:pcor}
\end{equation}
has been introduced.

The general structure of the variational trial state is that it is created
by an intersite projection operator ${\hat P}$ acting on a mean--field
reference state $|\Phi\rangle$. On--site correlations (such as local singlet
formation) are included in $|\Phi\rangle$, and intersite correlations are
controlled by ${\hat P}$. Both $|\Phi\rangle$ and ${\hat P}$ can contain
variational parameters.

Analogous trial states could be written down for higher--dimensional
lattices. Actually, these would be physically more acceptable: our variational
method is a {\sl correlated mean field} method. Thus the overall appearance
of our results is what we would expect for three--dimensional systems.
Subtle features particular to one dimension \cite{SM} are likely to be
missed by the present treatment. However, the algebra needed for the
variational method can be executed fully in one dimension while it would
involve further approximations in higher dimensions. Therefore we stick to
the one--dimensional case anticipating that care has to be exercised in
interpreting the findings.

The method we use has been introduced for spin chains by Virosztek \cite{V};
later it was applied to the one--dimensional Hubbard model by Penc and one
of us \cite{FP}.

We will be working with homogeneous states for which
\begin{equation}
\sum_{i} (S_{i}^{z} + \tau_{i}^{z}) = 0
\label{eq:sz0}
\end{equation}
The motivation comes from considering the single--site mean--field
solution \cite{Don} of the original necklace model $W>0$, $J>0$, $\rho=0$.
In the simplest approximation, the ground state is an $x$--$y$
antiferromagnet if $J<W$, and a collection of independent singlets if $J>W$,
with a continuous phase transition at $J=W$. The finding of a ground state
phase transition between a magnetic and a ``Kondo--like'' state has long
been the source of inspiration for continuing research in the Kondo lattice
physics. On  the other hand, it has been a matter of debate whether
this transition is an exact consequence of the model (\ref{eq:ham}), or an
artefact of the simple approximation. At the simplest level, one can point it
out that the description of the non--ordered state as strictly singlet is
just a zeroth--order approximation since for any finite $W/J$, however small,
the process
\begin{equation}
|1/2, -1/2\rangle_{1}|-1/2, 1/2\rangle_{2} \longrightarrow
|1/2, 1/2\rangle_{1}|-1/2, -1/2\rangle_{2}
\label{eq:hopi}
\end{equation}
will mix in local triplets.

%It is a common experience that while the effective field treatment can do
%quite well for the ordered state, it gives only a crude characterization of
%the disordered state. It is the latter aspect that we wish to improve.
It is natural to expect that the description in terms of local singlets
becomes correct only in the limit $J/W\to\infty$, and for finite $J$, there
will be spin--spin
correlations between the sites. It is an interesting question whether the
mean--field transition survives the inclusion of such correlations.

\section{Variational trial states for the necklace model}

\subsection{Homogeneous states}

In accordance with (\ref{eq:sz0}), in (\ref{eq:phii}) we choose
\begin{equation}
|\alpha_{1}| = |\alpha_{4}| = \alpha  \ \ \ \mbox{and}\ \ \
|\alpha_{2}| = |\alpha_{3}| = \beta
\end{equation}
One still has to decide the relative phase factors.

The ground state wave function can be chosen as real, which says all the
parameters are real. The remaining task is to specify the relative signs
of the $\alpha$'s. This we
do by requiring that non--diagonal processes should give a negative
contribution to the energy, whenever that is possible. The on--site Kondo
spin--flip term involves the factor $\alpha_{2}\alpha_{3}$; the term can be
made negative if
\begin{equation}
{\rm sg}(\alpha_{2}\alpha_{3})=-{\rm sg}(J)
\label{eq:sg1}
\end{equation}
As we are going to see, there are two kinds of contributions coming from the
intersite $\tau$--spin--flip processes. One of them, which acts between sites
with antiparallel $S$--spins, was illustrated in (\ref{eq:hopi}). Here all
four local states appear once, so the term comes with the factor
$\alpha_{1}\alpha_{2}\alpha_{3}\alpha_{4}$. The corresponding energy term
can be made negative if we prescribe
\begin{equation}{\rm sg}(\alpha_{1}\alpha_{4})={\rm sg}(J)\cdot{\rm sg}(W)
\label{eq:sg2}
\end{equation}
In the other kind of hopping process, the $S$--spins are parallel
\begin{equation}
|1/2, 1/2\rangle_{1}|1/2, -1/2\rangle_{2} \longrightarrow
|1/2, -1/2\rangle_{1}|1/2, 1/2\rangle_{2}
\end{equation}
In the corresponding term, all variational parameters are raised to even
powers, so the sign of the contribution is the same as the sign of $W$.
In particular, for $W>0$, a homogeneous Ansatz does not permit to gain
energy from this kind of process.

No difficulties arise if we stick to the case $\rho=0$: In (\ref{eq:uni3})
we have shown that $W>0$ and $W<0$ are equivalent, so we can choose $W<0$,
and have all contributions negative. Still, it is interesting to remember
that the equivalent solution of the $W>0$ problem is a two--sublattice
antiferromagnetic state which can be generated from the homogeneous Ansatz
by acting on it with the transformation ${\hat U}_{2}$ given in
(\ref{eq:uni2})
\begin{eqnarray}
\lefteqn{{\hat U}_{2}\cdot\prod_{i=1}^{L}  (\alpha|1\rangle_{i} -
\beta|2\rangle_{i} + \beta|3\rangle_{i} - \alpha|4\rangle_{i})}\nonumber \\
& = &  \prod_{i=1}^{L/2}  (\alpha|1\rangle_{2i-1} -
\beta|2\rangle_{2i-1} +
\beta|3\rangle_{2i-1} - \alpha|4\rangle_{2i-1})\cdot
\nonumber \\[3mm]
& &   (\alpha|1\rangle_{2i} +
\beta|2\rangle_{2i} - \beta|3\rangle_{2i} - \alpha|4\rangle_{2i})
\end{eqnarray}
The two--sublattice form is just Doniach's \cite{Don} Ansatz for the case
$W>0$.

Henceforth we keep $W<0$. Depending on the sign of $J$, there are two
different classes of wave functions. To be specific, we choose $J>0$.
The Ansatz for the ground state reads
\begin{equation}
|\Psi\rangle = \prod_{i} {\hat P}_{i,i+1}
(\alpha|1\rangle_{i} + \beta|2\rangle_{i} -
\beta|3\rangle_{i} + \alpha|4\rangle_{i})
\label{eq:ans1}
\end{equation}
\\
The structure of the correlator can be represented by the matrix
\begin{equation}
{\hat O} = \left( \begin{array}{cccc}
             O(\xi_{1},\xi_{1}) & O(\xi_{1},\xi_{2}) & O(\xi_{1},\xi_{3})&
             O(\xi_{1},\xi_{4}) \\
             O(\xi_{2},\xi_{1})  & O(\xi_{2},\xi_{2}) & O(\xi_{2},\xi_{3})  &
             O(\xi_{2},\xi_{4})  \\
             O(\xi_{3},\xi_{1}) & O(\xi_{3},\xi_{2}) & O(\xi_{3},\xi_{3})&
             O(\xi_{3},\xi_{4}) \\
             O(\xi_{4},\xi_{1})  & O(\xi_{4},\xi_{2}) & O(\xi_{4},\xi_{3})  &
             O(\xi_{4},\xi_{4})
             \end{array}
              \right )
              \label{eq:omat}
\end{equation}
\\
Considering the obvious symmetry
\begin{equation}
O(\xi_{i}, \xi_{j}) = O(\xi_{j}, \xi_{i})
\end{equation}
and further restrictions arising from $\uparrow$--$\downarrow$ symmetry,
still leaves us with more variational parameters than one could easily
handle. One has to try to guess what are the relevant correlations. We
return to this question later. Before doing that, we outline the general
formalism.

\subsection{Transfer matrix formalism}

We have to calculate the ground state energy
\begin{equation}
E = \frac{\langle\Psi|H|\Psi\rangle}{\langle\Psi|\Psi\rangle}
\label{eq:E}
\end{equation}
and minimize it with respect to the variational parameters.

Let us keep the matrix ${\hat O}$ and the choice of the $\alpha$'s as yet
unspecified and calculate the norm $\langle\Psi|\Psi\rangle$. Expanding
$|\Psi\rangle$ in the orthonormal basis formed as the direct product of the
single--site bases (\ref{eq:bas})
\begin{equation}
|\Psi\rangle = \sum_{\xi_{1}=1}^{4} \sum_{\xi_{2}}\ldots \sum_{\xi_{L}}
\prod_{i=1}^{L}\left(\alpha_{\xi_{i}} O(\xi_{i},\xi_{i+1})\right)
|\xi_{1}\rangle_{1}|\xi_{2}\rangle_{2}\ldots|\xi_{L}\rangle_{L}
\end{equation}
we find that the norm is given by a sum over configurations, which is similar
to the partition function of a one--dimensional classical lattice model. It is
advantageous to introduce the transfer matrix ${\hat T}$ as
\begin{equation}
T(\xi_{i},\xi_{i+1}) = |\alpha_{\xi_{i}}|O^{2}(\xi_{i},\xi_{i+1})
|\alpha_{\xi_{i+1}}|
\label{eq:tr1}
\end{equation}
whereupon the norm becomes
\begin{eqnarray}
\langle\Psi|\Psi\rangle & = & \sum_{\xi_{1}} \sum_{\xi_{2}}\ldots
\sum_{\xi_{L}} T(\xi_{1},\xi_{2})T(\xi_{2},\xi_{3})\ldots
T(\xi_{L-1},\xi_{L})T(\xi_{L},\xi_{1}) \nonumber \\
 & = & Tr({\hat T}^{L}) \longrightarrow x_{0}^{L}
\label{eq:norm}
\end{eqnarray}
where $x_0$ is the largest eigenvalue of the transfer matrix.

The next step is the calculation of expectation values. The knowledge of
$x_0$ suffices to determine the densities of quantities which are
diagonal in the representation (\ref{eq:bas}). The density of sites in
the state $|\xi\rangle$ is
\begin{equation}
n_{\xi} = \alpha_{\xi}^{2} \frac{\partial \ln{x_{0}}}{\partial
                                 \alpha_{\xi}^{2}}
\label{eq:nxi}
\end{equation}
The combined density of nearest--neighbour pairs in the configuration
$|\xi_{1}\rangle|\xi_{2}\rangle$, and its reverse
$|\xi_{1}\rangle|\xi_{2}\rangle$ is
\begin{equation}
n_{\xi_{1}\xi_{2}} = O^{2}(\xi_{1},\xi_{2}) \frac{\partial \ln{x_{0}}}
                                           {\partial O^{2}(\xi_{1},\xi_{2})}
\label{eq:n12}
\end{equation}
where the symmetry of the matrix ${\hat O}$ was exploited.

The straightforward analogy with classical statistical mechanics ceases when
we go over to the calculation of off--diagonal quantities such as
spin--flip amplitudes. At this stage, it becomes apparent that
we are dealing with a genuinely quantum--mechanical problem. We follow the
method introduced by Virosztek \cite{V}.

Let us illustrate the method on the example of the spin--flip part of the
Kondo term acting at site $m$
\begin{eqnarray}
\lefteqn{\langle\Psi | S^{+}_{m}\tau_{m}^{-} +
S^{-}_{m}\tau^{+}_{m}|\Psi\rangle =}\nonumber \\
& &
\sum_{\xi_{1}} \ldots \sum_{\xi_{m-1}}\sum_{\xi_{m}}\sum_{\xi_{m}^{'}}
\sum_{\xi_{m+1}} \ldots \sum_{\xi_{L}} \prod_{i=1}^{m-1}
\alpha_{\xi_{i}}^{2} \cdot \prod_{i=1}^{m-2}O^{2}(\xi_{i},\xi_{i+1})
\nonumber \\
& & \cdot \alpha_{\xi_{m}} O(\xi_{m-1},\xi_{m})O(\xi_{m},\xi_{m+1})
\langle\xi_{m}|S^{+}_{m}\tau_{m}^{-} + S^{-}_{m}\tau^{+}_{m}|\xi_{m}^{'}
\rangle \nonumber \\
& & \cdot\alpha_{\xi_{m}^{'}}
O(\xi_{m-1},\xi_{m}^{'}) O(\xi_{m}^{'},\xi_{m+1})
\cdot
\prod_{i=m+1}^{L}\left(\alpha_{\xi_{i}}^{2} O^{2}(\xi_{i},\xi_{i+1})\right)
\label{eq:ko1}
\end{eqnarray}
The expression is rather like the norm (\ref{eq:norm}) except that at site
$m$, a disturbance has occured which, via the correlators, influences the
sites $m-1$ and $m+1$. (\ref{eq:ko1}) is still the trace of a product
of matrices; however, in contrast to (\ref{eq:norm}), not all matrices
are ${\hat T}$. The sites $m-1$ and $m+1$ are connected by the matrix
${\hat K}_{\pm}$ rather than by ${\hat T}^{2}$
\begin{equation}
\langle\Psi | S^{+}_{m}\tau_{m}^{-} +S^{-}_{m}\tau^{+}_{m}|\Psi\rangle =
Tr\left( {\hat T}^{m-2}{\hat K}_{\pm}{\hat T}^{L-m}\right)
\end{equation}
Doing the sums over $\xi_{m}$ and $\xi_{m}^{'}$ we recall that the spin--flip
terms connect the states $|2\rangle$ and $|3\rangle$. It can be read off
from (\ref{eq:ko1}) that
\begin{eqnarray}
K_{\pm}(\xi_{m-1},\xi_{m+1}) & = & 2|\alpha_{\xi_{m-1}}\alpha_{\xi_{m+1}}|
\alpha_{2}\alpha_{3}\nonumber \\
& & \cdot O(\xi_{m-1},2)O(2,\xi_{m+1})O(\xi_{m-1},3)O(3,\xi_{m+1})
\label{eq:ko2}
\end{eqnarray}
The exponential dominance of the trace by factors of $x_0$ allows to deduce
\begin{equation}
\frac{\langle\Psi | S^{+}_{m}\tau_{m}^{-} +S^{-}_{m}\tau^{+}_{m}|\Psi\rangle}
{\langle\Psi|\Psi\rangle} = \frac{\langle X_{0}|{\hat K}_{\pm}|X_{0}\rangle}
{x_{0}^{2}}
\label{eq:ko3}
\end{equation}
where $|X_{0}\rangle$ is the eigenvector satisfying
${\hat T}|X_{0}\rangle = x_{0}|X_{0}\rangle$. Because of the symmetry of
${\hat T}$, it is of the form
\begin{equation}
|X_{0}\rangle = \frac{1}{\sqrt{2(1+k^{2})}}
                       \left( \begin{array}{c}
                       k \\
                       1 \\
                       1 \\
                       k
                       \end{array} \right)
\label{eq:X0}
\end{equation}

The intersite spin--flip process between sites $m$ and $m+1$
exerts an influence also on sites $m-1$ and $m+2$. This can be expressed by
a matrix ${\hat M}_{\pm}$ which is analogous to ${\hat K}_{\pm}$
\begin{eqnarray}
\lefteqn{M_{\pm}(\xi_{m-1},\xi_{m+2})  =  2\alpha^{2}\beta^{2}|
\alpha_{\xi_{m-1}}\alpha_{\xi_{m+2}}|} \nonumber \\
& & \cdot ( O(1,4)O(2,3)[O(\xi_{m-1},1)O(4,\xi_{m+2})O(\xi_{m-1},2)
O(3,\xi_{m+2})+\nonumber \\
& & O(\xi_{m-1},4)O(1,\xi_{m+2})O(\xi_{m-1},3)O(2,\xi_{m+2})]
 \nonumber \\
& & + O^{2}(1,3)O(\xi_{m-1},1)O(3,\xi_{m+2})O(\xi_{m-1},3)O(1,\xi_{m+2})+
\nonumber \\
& & O^{2}(2,4)O(\xi_{m-1},2)O(4,\xi_{m+2})O(\xi_{m-1},4)O(2,\xi_{m+2}))
\label{eq:kin1}
\end{eqnarray}
which has to be replaced into
\begin{equation}
\frac{\langle\Psi | \tau^{+}_{m}\tau_{m+1}^{-} +\tau^{-}_{m}\tau^{+}_{m+1}
|\Psi\rangle}{\langle\Psi |\Psi\rangle} =
\frac{\langle X_{0}|{\hat M}_{\pm}|X_{0}\rangle}{x_{0}^{3}}
\label{eq:kin2}
\end{equation}
Another useful quantity is the transverse spin polarization which turns out to
be the order parameter of the ground state. The matrix belonging to
$S_{m}^{x}$ is
\begin{eqnarray}
{\hat P}_{S}(\xi_{m-1},\xi_{m+1}) & = & 2\alpha\beta
|\alpha_{\xi_{m-1}}\alpha_{\xi_{m+1}}|
\nonumber \\
& & \cdot( O(\xi_{m-1},1)O(1,\xi_{m+1})O(\xi_{m-1},2)O(2,\xi_{m+1}) +
\nonumber \\
& & O(\xi_{m-1},3)O(3,\xi_{m+1})O(\xi_{m-1},4)O(4,\xi_{m+1}))
\label{eq:sxm}
\end{eqnarray}
while the matrix belonging to $\tau_{m}^{x}$ is the very similar
\begin{eqnarray}
{\hat P}_{\tau}(\xi_{m-1},\xi_{m+1}) & = & -2\alpha\beta
|\alpha_{\xi_{m-1}}\alpha_{\xi_{m+1}}|
\nonumber \\
& & \cdot( O(\xi_{m-1},1)O(1,\xi_{m+1})O(\xi_{m-1},3)O(3,\xi_{m+1}) +
\nonumber \\
& & O(\xi_{m-1},2)O(2,\xi_{m+1})O(\xi_{m-1},4)O(4,\xi_{m+1}))
\label{eq:sxt}
\end{eqnarray}
The sign difference in (\ref{eq:sxm}) and (\ref{eq:sxt}) shows that (as
expected for an antiferromagnetic Kondo interaction) the $S$ and $\tau$
polarizations point in opposite directions.

\subsection{The minimization procedure}

Now we take particular forms of the Ansatz and work out the consequences.
In the process, we hope to learn which correlations are most relevant for
getting a correct description in different regimes of the coupling constant
$J/W$.

\subsubsection{Simple spin correlations}

Since the mean field solution gives a magnetic--to--nonmagnetic ground state
phase transition, the first idea could be to incorporate short range
spin--spin correlations. Though only the $\tau$ spins are subject to
intersite interactions, their ordering induces a similar ordering of the
$S$--spins. (Since the Kondo coupling is antiferromagnetic, the
local $S$--moment is antiparallel to the local $\tau$--moment.) Therefore we
incorporate  in the Ansatz both $\tau$--$\tau$ and $S$--$S$ intersite
correlations, controlled by the independent variational parameters
$\lambda_{S}$ and $\lambda_{\tau}$. Each antiparallel $\tau$--$\tau$ pair
brings a factor $\lambda_{\tau}$, and each antiparallel $S$--$S$ pair a factor
$\lambda_{S}$. The corresponding matrix of the correlation coefficients
$O(\xi_{i},\xi_{i+1})$ can be
written as
\begin{equation}
{\hat O}_{1} = \left( \begin{array}{cccc}
              1 & \lambda_{S} & \lambda_{\tau} & \lambda_{S}\lambda_{\tau} \\
              \lambda_{S} & 1 & \lambda_{S}\lambda_{\tau} & \lambda_{\tau} \\
              \lambda_{\tau} & \lambda_{S}\lambda_{\tau} & 1 & \lambda_{S} \\
              \lambda_{S}\lambda_{\tau} & \lambda_{\tau} & \lambda_{S} & 1
              \end{array}
              \right )
              \label{eq:o2mat}
\end{equation}
In principle, we could have included intersite $S-\tau$ correlations as
well. Since the motivation for these is not immediately clear, and the
minimization difficult enough with the parameters we already have, we omit
them.

${\hat O}_{1}$ which describes fluctuating magnetism, should work reasonably
well in the weak--to--intermediate coupling regime where it means an
improvement over the mean--field finding of static long--range order.

The largest eigenvalue of the transfer matrix is
\begin{eqnarray}
x_{0} & = & \frac{1}{2}\left[(\alpha^{2}+\beta^{2})(1+
\lambda^{2}_{S}\lambda^{2}_{\tau})+\right. \nonumber \\
& & \left.\sqrt{(\alpha^{2}-\beta^{2})^{2}
(1+\lambda^{2}_{S}\lambda^{2}_{\tau})^{2}+
4\alpha^{2}\beta^{2}(\lambda_{S}^{2}+\lambda_{\tau}^{2})^{2}}
\right]
\label{eq:x0a}
\end{eqnarray}
and $k$ appearing in the corresponding eigenvector
\begin{eqnarray}
k & = & \frac{1}{2\alpha\beta(\lambda_{S}^{2}+\lambda_{\tau}^{2})}
\left[(\alpha^{2}-\beta^{2})^{2}(1+
 \lambda^{2}_{S}\lambda^{2}_{\tau})^{2}+\right. \nonumber \\[3mm]
& & \left. \sqrt{(\alpha^{2}-\beta^{2})^{2}
(1+\lambda^{2}_{S}\lambda^{2}_{\tau})^{2}+
4\alpha^{2}\beta^{2}(\lambda_{S}^{2}+\lambda_{\tau}^{2})^{2}}\right]
\label{eq:ka}
\end{eqnarray}
These quantities enter the various terms of the ground state energy.
Using $\uparrow$--$\downarrow$ symmetries, the $z$--$z$ part of the
Kondo term can be derived from (\ref{eq:nxi}) as
\begin{eqnarray}
\frac{\langle\Psi | S^{z}_{m}\tau_{m}^{z} |\Psi\rangle}
{\langle\Psi|\Psi\rangle} & = & \frac{(\alpha^{2}-\beta^{2})(1+
\lambda^{2}_{S}\lambda^{2}_{\tau})}{4\sqrt{(\alpha^{2}-\beta^{2})^{2}
(1+\lambda^{2}_{S}\lambda^{2}_{\tau})^{2}+
4\alpha^{2}\beta^{2}(\lambda_{S}^{2}+\lambda_{\tau}^{2})^{2}}}
\label{eq:ko4}
\end{eqnarray}
For the spin--flip part, we use (\ref{eq:ko3}) to arrive at
\begin{equation}
\frac{\langle\Psi | S^{+}_{m}\tau_{m}^{-} +S^{-}_{m}\tau^{+}_{m}|\Psi\rangle}
{\langle\Psi|\Psi\rangle} =
-\frac{2\beta^{2}\lambda^{2}_{S}\lambda^{2}_{\tau}(\alpha k+\beta)^{2}}
{x_{0}^{2}(1+k^{2})}
\label{eq:ko5}
\end{equation}
Finally, (\ref{eq:kin2}) is used to derive
\begin{equation}
-\frac{W}{2} \frac{\langle\Psi | \tau^{+}_{m}\tau_{m+1}^{-} +
\tau^{-}_{m}\tau^{+}_{m+1}|\Psi\rangle}
{\langle\Psi|\Psi\rangle} = - W\frac{\alpha^{2}\beta^{2}\lambda_{\tau}^{4}
(1+\lambda_{S}^{2})^{3}(\alpha k+\beta)^{2}}{x_{0}^{3}(1+k^{2})}
\label{eq:kin3}
\end{equation}
To get a feeling for the structure of the result, let us first study
the case $J=0$ when only the term (\ref{eq:kin3}) remains. Maximum freedom
for $\tau$--spin--flip is obtained for $\alpha=\beta=1/\sqrt{2}$ which leads
to an expression which is independent of $\lambda_{S}$:
$-2\lambda_{\tau}^{4}/(1+\lambda_{\tau})^{2}$. This has its minimum at
$\lambda_{\tau}=\sqrt{2}$. The minimum energy $-8W/27$ amounts to 94\% of
the exact value $-W/\pi$.

For small $J/W$, an expansion in terms of the small quantities
$\beta/\alpha-1$, and $t-\sqrt{2}$ yields
\begin{equation}
\langle H\rangle \approx -\frac{8}{27}W - \frac{2}{9}J - \frac{1}{36}
\frac{J^{2}}{W}
\end{equation}
which corresponds to $\lambda_{S}\approx 1$ and $\lambda_{\tau}\approx
\sqrt{2}(1-3J/16W)$.

For general $J/W$, minimization was carried out numerically. The result for
the ground state energy is shown in Fig. 1. In the interval $0\le
J/W<1$, the ground state energy shows an improvement over the simple
mean--field result, which is quite substantial in the small--$J$ regime.
However, as $J/W$ is increased slightly beyond 1, the solution seems to
gradually approach that obtained by the single--site mean field treatment.

We confirmed the existence of a sharp phase transition by a semi--analytic
argument by expanding the energy in terms of the small parameter
$\alpha/\beta$. Strictly for $\alpha=0$, the intersite hopping contribution
(\ref{eq:kin3}) vanishes, and the Kondo energy has its minimum value $-3J/4$
for $\lambda_{S}\lambda_{\tau}=1$. In the neighbourhood of the transition
we expect (and find) that $\delta=\lambda_{S}-1/\lambda_{\tau}$ is also
very small, so
we can make an additional expansion in terms of $\delta$. The leading terms
of the energy can be written as
\begin{equation}
\langle H \rangle \approx -\frac{3}{4}J +
[f_{1}(J/W,\lambda_{\tau})+f_{2}(J/W,\lambda_{\tau})\cdot\delta]
\left(\frac{\alpha}{\beta}\right)^{2}
\end{equation}
$f_2$ is negative but an energy lowering due to an infinitesimally small
$\delta$ is impossible if $f_1$ is a finite positive quantity. We found that
the minimum of $f_1$ changes sign at $J/W\approx 1.059$. For larger $J$'s, the
total energy increment in non--negative which requires $\alpha=0$, i.e., we
are back at the single--site mean field solution.

Below the threshold value, the ground state is ordered, having
non--vanishing expectation values of $\tau^x$ and $S^x$
\begin{equation}
\langle\tau^{x}\rangle = -\frac{\alpha\beta\lambda_{\tau}^{2}
(1+\lambda_{\tau}^{2})^{2}(\alpha k+\beta)^{2}}
{x_{0}^{2}(k^{2}+1)}
\end{equation}
\\[3mm]
\begin{equation}
\langle S^{x}\rangle = \frac{\alpha\beta\lambda_{\tau}^{2}
(1+\lambda_{S}^{2})^{2}(\alpha k+\beta)^{2}}
{x_{0}^{2}(k^{2}+1)}
\end{equation}

Our experience with the Ansatz specified by (\ref{eq:o2mat}) can be
summarized like this: introducing {\sl independent} $\tau$--$\tau$ and
$S$--$S$ correlations leads to considerable improvement in the description
of the ordered state. However, there is still a phase transition from
an ordered to a non--ordered ground state at the threshold value $J=1.059W$
which is near to $J=W$ ontained in the ordinary mean field treatment
(Fig. 2). For $J>1.059W$, the
description reduces to that obtained from the product trial state
(\ref{eq:phii}), i.e., it gives an array of decoupled singlets. The
description of the high--$J$ regime can be improved by postulating less
obvious kinds of intersite correlations.

\subsubsection{Composite spin correlations}

Simple spin correlations failed to provide an acceptable characterization
of the ground state for $J/W>1$. To understand the nature of this state,
let us remember that in the large--$J$ limit, the process (\ref{eq:hopi})
prevents the system  from freezing into a collection of singlets: it will
keep on creating antiparallel pairs of local triplets.
However, these pairs should then dissolve into singlets again,
otherwise a high--energy situation remains sustained. Thus there must be
a tendency for antiparallel components of local triplets to remain at
nearest neighbour distance, which we try to enforce by $O(1,4)=
O(4,1)=\eta$. If we wish, we can assist the pair
creation of local triplets by enhancing the antiparallel correlations in
nearby singlets via $O(2,3)=O(3,2)=\zeta$.
Thus we are led to consider
\begin{equation}
{\hat O}_{2} = \left( \begin{array}{cccc}
              1 & 1 & 1 & \eta \\
              1 & 1 & \zeta & 1 \\
              1 & \zeta & 1 & 1 \\
              \eta & 1 & 1 & 1
              \end{array}
              \right )
              \label{eq:o3mat}
\end{equation}
\\
It should be emphasized that here we are considering the correlations of
composite objects made up of $S$-- and $\tau$--spins. The correlations
we introduced do not factorize into independent $S$--$S$ and $\tau$--$\tau$
correlations.

The largest eigenvalue of the transfer matrix (\ref{eq:tr1}) is now
\begin{equation}
x_{0} = \frac{1}{2}\left( \alpha^{2}(1+\eta^{2})+\beta^{2}(1+\zeta^{2})+
\sqrt{[\alpha^{2}(1+\eta^{2})-\beta^{2}(1+\zeta^{2})]^{2}+
16\alpha^{2}\beta^{2}}
\right)
\label{eq:x0b}
\end{equation}
which belongs to an eigenvector of the form (\ref{eq:X0}) with
\begin{equation}
k = \frac{1}{4\alpha\beta}\left( \alpha^{2}(1+\eta^{2})-\beta^{2}(1+\zeta^{2})
+\sqrt{[\alpha^{2}(1+\eta^{2})-\beta^{2}(1+\zeta^{2})]^{2}+
16\alpha^{2}\beta^{2}}\right)
\label{eq:X0b}
\end{equation}
\\
The $z$--$z$ part of the Kondo coupling is
\begin{equation}
\frac{\langle\Psi | S^{z}_{m}\tau_{m}^{z} |\Psi\rangle}
{\langle\Psi|\Psi\rangle} = \frac{1}{4}\cdot
\frac{\alpha^{2}(1+\eta^{2})-\beta^{2}(1+\zeta^{2})}
{\sqrt{[\alpha^{2}(1+\eta^{2})-\beta^{2}(1+\zeta^{2})]^{2}+
16\alpha^{2}\beta^{2}}}
\label{eq:ko6}
\end{equation}
while the spin--flip part is found to be
\begin{equation}
\frac{\langle\Psi | S^{+}_{m}\tau_{m}^{-} +S^{-}_{m}\tau^{+}_{m}|\Psi\rangle}
{\langle\Psi|\Psi\rangle} = -\frac{2\beta^2 (\alpha k+\beta\zeta)^2}
{x_{0}^{2}(1+k^{2})^{2}}
\label{eq:ko7}
\end{equation}
The $\tau$--spin--flip energy is given by
\begin{equation}
-\frac{W}{2} \frac{\langle\Psi | \tau^{+}_{m}\tau_{m+1}^{-} +
\tau^{-}_{m}\tau^{+}_{m+1}|\Psi\rangle}
{\langle\Psi|\Psi\rangle} = -\frac{2\alpha^{2}\beta^2 (1+\eta\zeta)
[\alpha(1+\eta)k+\beta(1+\zeta)]^2}
{x_{0}^{3}(1+k^{2})^{2}}
\label{eq:kin4}
\end{equation}
The energy expression can be put together from (\ref{eq:ko4}),
(\ref{eq:ko5}), and (\ref{eq:kin3}). It has to be minimized with respect to
the three independent variational parameters $\alpha/\beta$, $\eta$, and
$\zeta$.

Optimization has to be done numerically and the results are shown in
Fig. 3 for the ground state energy and in Fig. 4 for (the absolute value of)
the order parameter
\begin{equation}
\langle S^{x} \rangle = -\langle \tau^{x} \rangle =
\frac{\alpha\beta[\beta(1+\zeta)+\alpha k(1+\eta)]^{2}}{x_{0}^{2}
(1+k^{2})}
\end{equation}
\\[3mm]
The most relevant feature is the suppression of the phase transition:
{\sl the ground state is ordered for all $J$}. The ordered moment is of
$O(1)$ for weak coupling and gets gradually suppressed in the strong
coupling regime. We find it intriguing that a hint of reduced moment
magnetism emerges in a model which is thought to correspond to the
single--channel Kondo lattice.

Details of the behaviour can be discussed in limiting cases:

For $J=0$, only the $x$--$y$ term remains.
Due to the symmetrical role played by local singlets and triplets in enhancing
the mobility of the $\tau$--spins the
minimum corresponds to $\alpha=\beta$, and $\eta=\zeta$.
The lowest value is $\sim -0.281W$ which we find at $\eta
\approx 1.512$. Thus even in the limit which is the exact opposite
of what the trial state is intended for, a significant improvement over the
mean field solution (energy $-0.25W$ for $\eta=\zeta=1$, $\alpha=\beta$) is
achieved. The ordered moment is $\langle\tau^{x}\rangle\approx 0.452$.

For large $J$, the density of local triplets has to be small, meaning
$\alpha\ll 1$, while the few triplets that are left must occur in
antiparallel pairs so $\eta$ must become large. In contrast, the correlations
governed by $\zeta$ become relatively unimportant; for the sake of the
present argument, we set $\zeta=1$. Introducing the convenient
parametrization $\alpha=\cos{\varphi}$, $\beta=\sin{\varphi}$,
$\Delta\varphi=\pi/2-\varphi$ is one of the small parameters;
the other is $1/\eta$. Expanding in these, the structure of the energy
expression suggests to look for the asymptotic solution in the form
\begin{equation}
\eta = \frac{c_{1}W}{J(\Delta\varphi)^{2}}
\end{equation}
and
\begin{equation}
\Delta\varphi=\frac{c_{2}W}{J}
\end{equation}
where $c_1$ and $c_2$ are to be determined from optimizing the leading
contributions to the energy
\begin{equation}
\langle H\rangle \approx -\frac{3}{4}J -\frac{W^2}{J}
\left(\frac{c_{1}}{2}-c_{1}^{2}
\right) + \frac{3c_{1}^{4}}{4(\Delta\varphi)^{2}}\frac{W^{4}}{J^{3}} +
J(\Delta\varphi)^{2}
\end{equation}
yielding
\begin{equation}
c_{1} = \frac{\sqrt{3}-1}{8}
\end{equation}
and
\begin{equation}
c_{2} = \sqrt{\frac{\sqrt{3}}{2}}c_{1}
\end{equation}
The leading terms of the energy become
\begin{equation}
\langle H\rangle \approx -\frac{3}{4}J -\frac{\sqrt{3}-1}{32}\cdot
\frac{W^2}{J}
\label{eq:enasy}
\end{equation}
which is obtained for
\begin{equation}
\eta \approx 8\left( 1+\frac{1}{\sqrt{3}} \right)\cdot \frac{J}{W}
\end{equation}
and
\begin{equation}
\alpha\sim \frac{W}{J}
\end{equation}
\\[3mm]
The asymptotic behaviour of the order parameter is found to be
\begin{equation}
\langle\tau^{x}\rangle \approx \frac{3^{1/4}(\sqrt{3}-1)}{8\sqrt{2}}\cdot
\frac{W}{J}\cdot\left[ 1+ \frac{(\sqrt{3}-1)}{8}\cdot\frac{W}{J}\right]
\label{eq:txasy}
\end{equation}
Thus our variational method recovered the correct order of magnitude
$\sim -W^{2}/J$ of the ground state energy in the large--$J$ limit: it is
what we would expect from perturbation theory. The nature of the problem is
similar to that of the large--$U$ behaviour of the Hubbard model where
nearest--neighbour holon--doublon correlations were found to be important
\cite{FP}.

In retrospect we can identify the reason why simple spin correlations are
insufficient in the large--$J$ regime. The correlation matrix
(\ref{eq:o2mat}) enforces $O(1,4)=O(2,3)$ while with (\ref{eq:o3mat})
we have found that $O(1,4)/O(2,3)$ has to become very large as $J/W$ increases.

\section{Discussion and conclusion}

We were trying to achieve a more detailed understanding of the behaviour of
the Kondo necklace model, with the eventual aim of finding results which may
be relevant to the physics of heavy fermion systems. We were considering
the necklace hamiltonian (\ref{eq:ham}) with $\rho=0$, i.e., the simplest
form introduced by Doniach \cite{Don}.

The single--site mean field solution of (\ref{eq:ham}) indicates that with
increasing $W/J$, a magnetic--to--Kondo (ground state) phase transition is
taking place. While this is a physically appealing result, the obvious
shortcomings of the characterization of the high--$J$ state as an array of
disconnected singlets may lead to worries that the phase transition is
merely an artefact of the approximation.

We improved the variational description of the ground state by allowing for
the presence of a variety of nearest--neighbour correlations. We carried out
the optimization in the one--dimensional case where the transfer matrix
method \cite{V} can be used to calculate the relevant expectation values.

In Section 3.3.1 we used the simple spin correlations whose presence can be
inferred from the ordering. These led to a better ground state energy in the
small--to--intermediate $J$ regime, and pushed the phase transition point
slightly upwards (Figures 1 and 2). However, the high--$J$ state remained
as structureless as in the simplest mean field approximation.

In Section 3.3.2 we learned that the physically interesting ones are the
composite spin correlations which could be expressed as expectation values
of products of four spin--operators, involving both kinds of spins (we could
choose $\langle S_{i}^{+}\tau_{i}^{+}S_{i+1}^{-}\tau_{i+1}^{-}\rangle$).
The relevance of these could be guessed from perturbation theory: they
describe that for $J\gg W$, the ground state is almost singlet, with a
sprinkling of a few nearest--neighbour antiparallel triplet pairs. The
inclusion of these correlations suppresses the phase transition completely
(Figures 3 and 4), the ground state remains ordered for any finite $J$.
Furthermore, the ground state energy, and the concentration of local triplets
have the order of magnitude expected from perturbation theory. The tail
regime
of the order parameter (Fig. 4) is a tantalizing hint that small ordered
moments may be a part of the physics of the orbitally non--degenerate Kondo
lattice.

However, we have to be extremely cautious about the conclusions to be drawn
from our results. After all, for the one--dimensional Kondo necklace,
powerful techniques have provided a number of essentially exact statements,
and these tend to be in disagreement with our findings. General arguments
suggest \cite{SM} that for almost the entire range of $J/W$ values, the
excitation spectrum is gapped but at some small $J/W$, the possibility of a
phase transition can not be excluded. In fact, exact diagonalization studies
\cite{SaSo} revealed the existence of a Kosterlitz--Thouless type transition
from the gapped high--$J$ state (with exponential decay of spin correlations)
to a gapless low--$J$ state (with algebraic decay of spin correlations). In
any case, the ground state never has true long--range order.

This point in itself should not be too worrying. Our correlated mean field
approach is supposed to work where long--range order is in principle possible,
i.e., foremost in three dimensions. It is just technical convenience which
made us to stick to one dimension but we could argue that the general
appearance of our results is the same as what a much more cumbersome
three--dimensional evaluation should give. This would still allow us to hope
that the three-dimensional pseudospin model would have a ground state with
small ordered moments.

We have to be, however, aware of a subtler kind of difficulty as well.
Prescribing an Ansatz means that the system is permitted to seek a low--energy
state in a certain manner. This can lead to a good estimate for the ground
state energy
(as it undoubtedly does) but does not necessarily imply that this is the
natural way how the energy gain in question arises. To cite an example, a
similar study of the Hubbard model \cite {FP} gave the correct order of
magnitude $-t^{2}/U$ for the energy at $U\gg t$ but ascribed it to a metallic
ground state which is patently false. Hence we should be warned that finding
the (functionally) correct asymptotic form (\ref{eq:enasy}) of the ground
state energy does not prove that the result (\ref{eq:txasy}) about the
long--range order is basically right. In fact, preliminary results
\cite{KKF} obtained from the Oguchi approximation indicate that the order
parameter $\langle\tau^{x}\rangle$ vanishes above a critical $J/W$. It
remains an outstanding question whether the three--dimensional Doniach
pseudospin model can support reduced moment magnetism.

\vspace{.8cm}

\centerline{ACKNOWLEDGEMENTS}
\vspace{.4cm}
The authors wish to express their gratitude to the International Centre
for Theoretical Physics for financial support, hospitality, and an
encouraging scientific atmosphere. P.F. is indebted also
to SISSA (Trieste) for the hospitality extended to him.

\newpage
\addcontentsline{toc}{subsubsection}{Figure captions}

{\bf\large Figure captions}
\\[1cm]
Fig. 1. Ground state energy (in units of $W$) versus $J/W$ for a trial
state with intersite $S$--$S$ and $\tau$--$\tau$ spin correlations. Thick
line: correlated mean field solution (transition at $J/W\approx 1.059$),
thin line: simple mean field (transition at $J/W=1$).
\\[3mm]
Fig. 2. The order parameter $\langle\tau^{x}\rangle$, for the same case as
in Fig. 2.
\\[3mm]
Fig. 3. Ground state energy (in units of $W$) for a trial state with
composite intersite correlations, belonging to the antiparallel alignment
of fluctuating triplets. Thick line: correlated mean field, thin line:
simple mean field.
\\[3mm]
Fig. 4. Order parameter $\langle\tau^{x}\rangle$ versus $J/W$ for
the same case as Fig. 3. The ground state phase transition found in the
simple mean field solution (thin line) is suppressed, according to the
correlated mean field approach (thick line), the asymptotic behaviour is
$\sim W/J$.

\end{document}